\PassOptionsToPackage{x11names,dvipsnames,table}{xcolor}
\documentclass[acmsmall,screen,nonacm]{acmart}
\usepackage{listings}
\usepackage[capitalise]{cleveref}
\usepackage[most]{tcolorbox}
\usepackage{adjustbox}
\usepackage{todonotes}

\AtBeginDocument{}

\lstdefinelanguage{Ruby}{
  morekeywords={return, channels, with, stateful, stateless, services, def, end, import, as, if, for, else, elsif, while, in, do, class, module, require, include, then},
  sensitive=true,
  morecomment=[l]{\#},
  morestring=[b]',
  morestring=[b]"
}

\let\origthelstnumber\thelstnumber
\makeatletter
\newcommand*\Suppressnumber{%
\lst@AddToHook{OnNewLine}{%
\let\thelstnumber\relax%
\advance\c@lstnumber-\@ne\relax%
}%
}
\newcommand*\Reactivatenumber[1]{%
\setcounter{lstnumber}{\numexpr#1-1\relax}%
\lst@AddToHook{OnNewLine}{%
\let\thelstnumber\origthelstnumber%
\refstepcounter{lstnumber}%
}%
}
\makeatother

\newcommand{\role}[1]{{\color{purple}\texttt{#1}}}
\newcommand{\at}[1]{{\role{@#1}}}
\newcommand{\hl}[1]{{\color{blue!85!black}#1}}
\newcommand{\hlcode}[1]{\hl{\texttt{#1}}}
\newcommand{\fw}{\hl{\blacktriangleright}}
\newcommand{\lang}{\textsf{FaaSChal}}
\newcommand{\code}[1]{\cd{#1}}
\newcommand{\cd}[1]{\lstinline[keepspaces=true,mathescape=true]|#1|}
\newtcolorbox{codebox}[1][]{%
  colback=yellow!10!white,
  colframe=gray,
  width=#1,
  left=0pt,
  top=0pt,
  bottom=0pt,
  right=0pt
}

\setcopyright{rightsretained}
\copyrightyear{2024}
\acmYear{2024}

\makeatletter
\let\@authorsaddresses\@empty
\makeatother

\acmConference[CP '24]{Choreographic Programming}{Copenhagen}{Denmark}

\begin{document}

\title{Towards a Function-as-a-Service Choreographic Programming Language: Examples and Applications}

\author{Giuseppe De Palma}
\affiliation{
  \institution{Università di Bologna}
  \country{Italy}
  \and
  \institution{OLAS team, INRIA}
  \country{France}
}

\author{Saverio Giallorenzo}
\affiliation{
  \institution{Università di Bologna}
  \country{Italy}
  \and
  \institution{OLAS team, INRIA}
  \country{France}
}

\author{Jacopo Mauro}
\affiliation{
  \institution{University of Southern Denmark}
  \country{Denmark}
}

\author{Matteo Trentin}
\affiliation{
  \institution{Università di Bologna}
  \country{Italy}
  \and
  \institution{OLAS team INRIA}
  \country{France}
  \and
  \institution{University of Southern Denmark}
  \country{Denmark}
}

\author{Gianluigi Zavattaro}
\affiliation{
  \institution{Università di Bologna}
  \country{Italy}
  \and
  \institution{OLAS team, INRIA}
  \country{France}
}

\maketitle

\section{Introduction}

Choreographic Programming (CP) is a language paradigm whereby software
artefacts, called choreographies, specify the behaviour of communicating
participants. Choreographic programming is famous for its
correctness-by-construction approach to the development of concurrent,
distributed systems. In this paper, we illustrate \lang{}, a proposal for a CP
language tailored for the case of serverless Function-as-a-Service (FaaS). In
FaaS, developers define a distributed architecture as a collection of stateless
functions, leaving to the serverless platform the management of deployment and
scaling~\cite{ES19}. We provide a first account of a CP language tailored for
the FaaS case via examples that present some of its relevant features, including
projection. In addition, we showcase a novel application of CP. We use the
choreography as a source to extract information on the infrastructural relations
among functions so that we can synthesise policies that strive to minimise their
latency while guaranteeing the respect of user-defined constraints.

\section{Background on Serverless}

We start with a brief overview of serverless computing and the platforms
that support it.

Developers make a serverless application out of software units called functions,
which run in short-lived environments triggered by different kinds of events.
When an event such as an HTTP request, database change, file upload or scheduled
trigger occurs, the FaaS platform runs an instance of the function(s) liked to
that event. The platform runs the code after initialising an execution
environment, which is a secure and isolated context that provides all the
resources needed for the function lifecycle, typically implemented with virtual
machines and containers.

\begin{figure}[t]
\centering
\includegraphics[width=\textwidth]{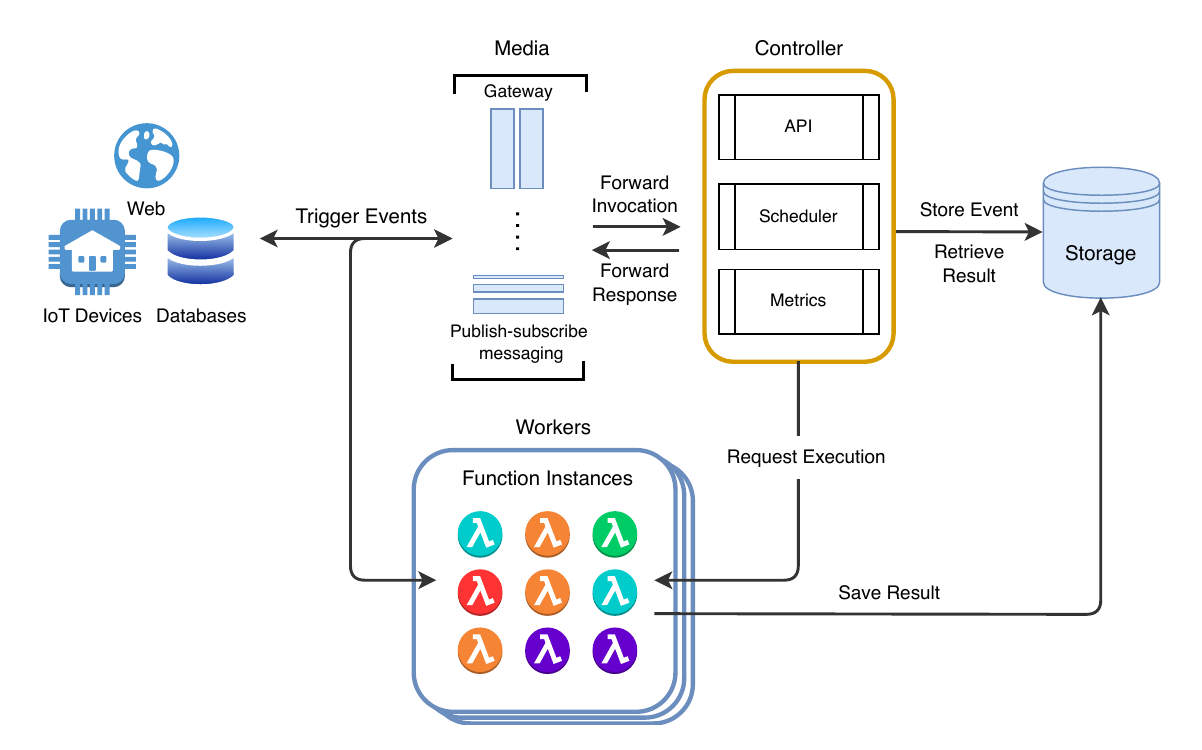}
\caption{A typical serverless platform architecture.}
\label{fig:serverless-architecture}
\end{figure}

We use \Cref{fig:serverless-architecture}, which depicts a typical serverless
platform architecture, to briefly introduce the components and processes behind
the execution of serverless functions, useful to contextualise our contribution.
The main components of a serverless platform, as shown in
\Cref{fig:serverless-architecture},  are the Controller and the Workers. The
Controller can receive requests to execute functions from various media, e.g.,
an HTTP gateway or a publish-subscribe messaging service like the Simple
Notification Service (SNS), by AWS. These media expose endpoints which
entities---like Web applications, IoT devices, and databases, as well as running
serverless functions---can trigger to invoke the execution of a function. The
Controller handles the allocation of functions on Workers based on the latter's
status (given a set of metrics like CPU and memory usage, collected by the
Controller). The Controller also handles the storage/retrieval of the
invocation/result from/to the caller.
In particular, the scheduler determines which Worker should execute an invoked
function based on factors such as current load, function requirements, and
resource availability. Once it receives the request to execute a function, the
Worker creates a new instance of that function, handling its execution
environment lifecycle, including provisioning, scaling, and teardown. 

\section{\lang{} by Example}

We introduce \lang{} with the example shown in \cref{lst:choreo}, where a
function orchestrates (a simplified version of) the training of an image AI
model. The whole routine is started by a \role{user}, who defines the queries to
extract from some databases the labels and images used in the training. The user
sends to a serverless function, called \role f, the queries, which \role f uses
to access two separate databases and obtain the images and labels (ordered and
paired). The function then launches the training of each pair image-label in a
separate function, called \role g, which finally triggers a third function,
called \role h, that acts as merger/integrator of the trained weights of the
model into a third database. The training process is asynchronous, i.e., the
user receives a response from the orchestrator function as soon as it terminates
the launching of the training of all image-label pairs.

\begin{figure}[t]
\begin{center}
\begin{codebox}[.85\textwidth]
\begin{lstlisting}[mathescape=true,basicstyle=\small\ttfamily,label=,caption={}]
stateful: $\role{user}$
stateless: $\role{f}${ Gateway }, $\role{g}${ SNS:"aws:sns" }, $\role{h}${ SNS:"aws:sns" }
services: DB1{ getData }, DB2{ getData }, DB3{ storeData }
import Collections::zip as zip$\at f$
import Model::fit as fit$\at g$
import Model::integrate as int$\at h$$\Suppressnumber$
$\Reactivatenumber{6}$
def main( queries$\at{user}$ )
  queries$\at{user}$ <-Gateway-> $\role f$ do | queries$\at f$ |
    queries$\at f$.labels <-> DB1: getData $\fw$ labels$\at f$
    queries$\at f$.images <-> DB2: getData $\fw$ images$\at f$
    for pair$\at f$ in zip(labels$\at f$, images$\at f$) do
      pair$\at f$ 
        $\fw$ -SNS-> $\role g$ $\fw$ fit 
        $\fw$ -SNS-> $\role h$ $\fw$ int 
        $\fw$ -> DB3: storeData
    end 
  end with "All training jobs started"$\at f$
end
\end{lstlisting}
\end{codebox}
\end{center}
\caption{The choreography of the serverless AI training program.}
\label{lst:choreo}
\end{figure}

In the choreography, we distinguish three main kinds of entities, found in the
preamble at lines 1--3. We start commenting on them from line 3 upwards. The
first kind is that of \cd{services}, which are passive entitites that interact
in the choreography providing labelled inbound request-response
operations---enumerated within curly brackets in the preamble. For example, in
\cref{lst:choreo}, the service \cd{DB1} offers a request-response operation
labelled \cd{getData}. In general, the caller can discard the response in a
request-response interaction, consuming the operation in a one-way fashion.
Going up, we find FaaS \cd{stateless} functions which: \emph{a}) must be
triggered/started by some other active entity via their media endpoints,
declared within brackets (e.g., \cd{Gateway}, \cd{SNS} and the trigger endpoint
annotations), \emph{b}) provide a request-response triggering behaviour (which
the triggerer can invoke in a one-way fashion, discarding their response),
\emph{c}) after their triggering, they cannot receive other messages (but they
can send outbound requests, in both one-way and request-response fashion). The
last kind of entity is that of \cd{stateful} participants, which are traditional
active processes (no triggering) that can interact with the other entities.

In the choreography, we can import operations (e.g., from libraries), as
showcased at lines 4--6, where the stateless functions \role{f}, \role{g}, and
\role{h} resp.\@ \cd{import} the operations \cd{zip} from the \cd{Collections}
library, \cd{fit} and \cd{int}(\cd{egrate}) from the \cd{Model} one. Since the
\cd{import} instruction has a target function (e.g., \role{f}), that
functionality is available/imported only at/by that function.

The statement we find at line 7 (closing at line 16) is a request-response from
the \role{user} to the stateless function \role{f}, of the form

\begin{lstlisting}[mathescape=true,numbers=none]
exp1$\at{role1}$ <- MEDIUM -> $\role{role2}$ do [| opt_var$\at{role2}$ |] ... end [ with exp2$\at{role2}$ ]
\end{lstlisting}

From left to right, we evaluate the expression \cd{exp} at \at{role1} and send
its value via the \cd{MEDIUM} the function is available at (e.g., the
\cd{Gateway} at line 7 of \cref{lst:choreo}) to trigger the execution of
function \role{role2}. This function can optionally bind the data sent from
\role{role1} to a local variable (\cd{opt_var}) and execute the code within the
block until its closure (\cd{end}). The function sends back a response, which is
empty unless specified through the suffix of the closure with the clause
\cd{with} followed by an expression at that function (\cd{exp$\at{role2}$})
evaluated to return a response---considering the body of the triggering block
and the initial binding within its scope.

Within the body of the block, we first find the request-response invocations
(\cd{<->}) to the respective operations \cd{getData} of \cd{DB1} and \cd{DB2} to
retrieve the data (resp.\@ \cd{labels} and \cd{images}) by
\role{f}.\footnote{The operation \cd{getData} in the example is blocking and,
thus, the second call to \cd{DB2} waits until the completion of the previous
call (which might take a long time) to proceed. To increase efficiency, a simple
extension of the language can include a parallel operator, like the one found in
AIOCJ~\cite{DGGLM17}, to send the two \cd{getData} calls in parallel, realising
a join pattern.}

To bind to a variable the value received by \role{f} as the response to the
request to the databases (both for \cd{DB1} and \cd{DB2}), we use the forward
operator \(\fw\). The idea behind \(\fw\), inspired by Choral~\cite{GMP24}, is
to naturally support a left-to-right reading of the interactions in a
choreography. Without \(\fw\), one would need to write an assignment like
\code{var$\at b$ = data$\at{a}$ <- MEDIUM -> $\ \at{b}$}, forcing the user to
first parse the expression on the right\footnote{Left-to-right: take the
\cd{data} from \role a, send it via \cd{MEDIUM}, instantiate \role b and return
the data sent by \role a to \role b as the expression's result.} and then go
back to the assignment of the resulting value to the
\lstinline|var|iable on the left.

Like in Choral, users can call unary functions with \(\fw\) in a point-free
style---i.e., \code{exp $\ \fw$ f1 $\ \fw$ f2} is syntactic sugar for \code{f2(f1(exp))}---which we extend to also work as a variable assignment operator.

At lines 11--15, after the retrieval of the \cd{labels} and \cd{images}, \role f
\cd{zip}s them together and, \cd{for} each \cd{pair}, it triggers a new instance
of the function \role g, sending to it the pair. Note that the triggering of
\role g is one-way, (represented by the communication \cd{-MEDIUM->} ), which
allows us to adopt the lightweight notation found, e.g., at line 12, instead of
the more complex one for request-responses we commented for \role f, above. At
triggering/reception, \role g performs the training (via the \cd{fit} operation)
and then triggers function \role h to \cd{int}egrate the data into \cd{DB3}
(invoking \cd{storeData} as a one-way operation).

A notable characteristic of \lang{} is that there is no need for coordination in
constructs such as loops and conditionals when the interaction concerns only
\cd{stateless} functions. Indeed, choreographic languages where processes are
stateful (and usually engage in a kind of session-oriented interaction) need
either the enforcement of knowledge of choice or amendments such as auxiliary
communications to ensure the causality/connectedness of the actions among the
processes~\cite{CDP11,LMZ13,DGGLM17,GMP24}. When conditionals/loops concern only
stateless functions (which, once triggered, cannot engage in further
synchronisations except for outbound request-responses) these issues do not
arise. As a consequence of this triggering behaviour for \cd{stateless}
functions, at each loop at lines 11--15, we bind the identifiers \role g and
\role h to resp.\@ new function instances, i.e., at each loop \role f requests
the instantiation of new copies of said functions. To further clarify the
relationship between knowledge of choice and stateless functions, consider the
example below

\begin{center}
  \code{if exp}\at f \code{then } \role f \code{-SNS->} \role g \code{else } \role f \code{-SNS->} \role h
\end{center}

If \role g and \role h were stateful processes, we would need to inform them
both on which direction the choreography shall proceed, according to the choice
taken by \role f (resulting from the evaluation of \code{exp}). Lacking this
piece of coordination, depending on the choice made by \role f, either \role g
or \role h would wait for \role f's call indefinitely (since either of them does
not know that \role f selected the other branch), exposing the program to
deadlocks. On the contrary, since all roles are \code{stateless} functions,
there is no need to inform, e.g., \role g that \role f is choosing the
\code{else} branch, because \role g is not running (it is triggered by \role f's
call) and has no risk of ending up in a deadlock state.

\section{Projection}

We show one of the typical applications of choreographic programming, which is
the generation of local code that implements the semantics of the source
choreography. In particular, this section aims to provide code examples that
FaaS developers can use to get a better grasp of the semantics of the example
in \cref{lst:choreo}. In the following, we use pseudocode inspired by Ruby and
Python and annotate the code to indicate to which entity it corresponds and
other information useful for deployment, e.g., the name that the FaaS platform
shall bind to the function and how it shall expose the function for consumption
(its \cd{MEDIUM}).

\begin{figure}[t]
\begin{small}
\begin{center}
\noindent\begin{minipage}{.49\textwidth}
\begin{codebox}[.96\textwidth]
\begin{adjustbox}{width=.75\textwidth}
\begin{lstlisting}[mathescape=true]
# user code
def main( queries )
  Gateway.invoke( "f", queries )
end
\end{lstlisting}
\end{adjustbox}
\end{codebox}
\end{minipage}
\begin{minipage}{.49\textwidth}
\begin{codebox}[\textwidth]
\begin{adjustbox}{width=.8\textwidth}
\begin{lstlisting}[mathescape=true]
# deploy as f, trigger GATEWAY
def main( queries )
 labels = DB1.getData( queries.labels )
 images = DB2.getData( queries.images )
 for pair in labels.zip( images ) do
  triggerFn( "g", "aws:sns", pair )
 end
 return { code: 200, body: "done" }
end
\end{lstlisting}
\end{adjustbox}
\end{codebox}
\end{minipage}

\noindent\begin{minipage}{.49\textwidth}
\begin{codebox}[.96\textwidth]
\begin{adjustbox}{width=\textwidth}
\begin{lstlisting}[mathescape=true]
# deploy as g, trigger SNS
import Model::fit as fit

def main( _ )
 return triggerFn( "h", "aws:sns", fit( _ ) )
end
\end{lstlisting}
\end{adjustbox}
\end{codebox}
\end{minipage}
\begin{minipage}{.49\textwidth}
\begin{codebox}[\textwidth]
\begin{adjustbox}{width=.7\textwidth}
\begin{lstlisting}[mathescape=true]
# deploy as h, trigger SNS
import Model::integrate as int

def main( _ )
  return DB3.storeData( int( _ ) )
end
\end{lstlisting}
\end{adjustbox}
\end{codebox}
\end{minipage}
\end{center}
\caption{Projections of the example from \cref{lst:choreo}.}
\label{lst:projection}
\end{small}
\end{figure}

We remind that \cd{services} are passive entities which the other participants
use as always-available operations, thus, the produced local code does not
include the sources for \cd{DB1}, \cd{DB2}, and \cd{DB3}.

Without going too much into the details of the pseudocode, we notice the most
salient features linked to serverless function programming. First, in the code
of functions, note the presence of a \cd{main} procedure, which is the one
canonically invoked by the platform to execute the behaviour of the function.
Second, we find the automatic injection of FaaS platform auxiliary
functionalities (one can make these functionalities platform-agnostic by
providing different implementations of the same API parametrised w.r.t.\@
specific deployments) provided to trigger the functions, e.g.,
\cd{Gateway.invoke} and \cd{triggerFn} resp.\@ found in the \role{user}'s and
functions' code. To keep the example lightweight, we did not introduce distinct
syntaxes for one-ways and request-responses---e.g., one can provide the same API
parametrised to either send a request and wait for a response to return it or
return immediately.

\section{Applications: The Case of Function Scheduling Policies}

The scheduling of functions, i.e., the allocation of functions over the
available workers, can substantially influence their performance. 
Indeed, effects like \emph{code locality}~\cite{HSHVAA16}---due to latencies in
loading function code and runtimes---or \emph{session
locality}~\cite{HSHVAA16}---due to the need to authenticate and open new
sessions to interact with other services---can substantially increase the run
time of functions.
Usually, serverless platforms implement opinionated policies that favour some
performance principle tailored for one or more of these locality principles.
Besides performance, functions can have functional requirements that the
scheduler shall consider. For example, users might want to ward off allocating
their functions alongside ``untrusted'' ones---common threat vectors in
serverless are limited function isolation and the ability of functions to
(surreptitiously) gather weaponisable information on the runtime, the
infrastructure, and the other tenants.

Although one can mix different principles to expand the profile coverage of a
given platform-wide scheduler policy, the latter hardly suits all kinds of
scenarios. This shortcoming motivated De Palma et
al.~\cite{PGMZ20,DGMTZ22,DGMTZ23a,DGMTZ23b} to introduce a YAML-like
declarative language used to specify scheduling policies to govern the
allocation of serverless functions on the nodes that make up a cluster, called
\emph{Allocation Priority Policies} (APP). Thanks to APP, the same platform can
support different scheduling policies, each tailored to meet the specific needs
of a set of related functions.

As an example of an application of \lang{}, we introduce a variant of APP. We
extract locality principles that emerge from the choreography---e.g., the loop
where \role{f} spawns many \role{g}s and \role{h}s presents a locality linked to
the time it takes \role{f} to contact \cd{SNS} and issue the call. Then, given a
description of the infrastructure topology and possible user-defined constraints
on the allocation of functions, we synthesise an APP script that strives to
orient the scheduling of functions to minimise their latency of execution, while
guaranteeing the respect of the constraints imposed by the user.

\subsection{The APP Language}
To define function-specific policies, APP assumes the association of each
function with a tag. In our examples, we directly use the function's reference
name as the tag, but the relation can be one-to-many to specify a policy shared
among a set of functions. Then, APP associates a tag to a policy, so that, at
runtime, the scheduler of the platform can pair each function with its APP
policy and follow the latter's scheduling logic.

In the APP variant we showcase, we assume to have the nodes of the cluster
associated with a label, i.e., several nodes can share the same label, e.g.,
\cd{group1}. In an APP script, users can specify a sequence of blocks (each
identified by YAML's list unit \cd{-}) associated with a tag. Each block
indicates on which nodes the scheduler can allocate the function. At function
invocation, the scheduler tries to allocate the function following the logic in
the first block, passing to the next only if none of the machines specified in
that block can host the function, and so on (exhausting all blocks causes the
invocation's failure). In APP, these nodes take the name of \hlcode{workers},
which is also the keyword used in the scripts to specify the label of the nodes
for that block. Besides \hl{workers}, APP lets users specify the strategy the
scheduler shall use to select among the indicated workers (e.g., choose at
random, for load-balancing) and when a worker becomes invalid (e.g., setting a
maximal threshold of concurrent functions running on it). The variant we present
does not use these options---but it is valid APP code nonetheless, since, when
omitted, APP uses default strategy and invalidation rules. The only additional
element in our variant's syntax is that of \hlcode{affinity}. This option
accepts a list of tags, where each tag can be prefixed by a \cd{!}. For
instance, if we have a tag \cd{a} with \hlcode{affinity}\cd{: b, !c}, it means
that function \cd{a} is affine with \cd{b}-tagged functions and anti-affine with
\cd{c}-tagged ones. Schedule-wise, ``anti-affine'' means that we cannot allocate
the function we want to schedule on a worker that contains instances of any of
its anti-affine functions---from the example, we cannot allocate an instance of
\cd{a} on workers hosting instances of \cd{c}. Complementarily, ``affine'' means
that we can allocate the function under scheduling only on workers that host at
least one instance of each of its affine functions---from
the example, we can allocate an instance of \cd{a} only on workers which have at
least one instance of \cd{b} running on them. These (anti-)affinity constraints
are useful to specify e.g., security concerns (like separating the execution of
trusted from untrusted functions to avoid possible security risks) and
performance (like the allocation of functions on the same worker to let them
reuse a pool of connections to a database). In APP, \hlcode{affinity}
constraints are not symmetric, i.e., if we set \cd{f} affine with \cd{g} it does
not imply that \cd{g} is affine with \cd{f} (but one can symmetrise the relation
by adding the complementary constraint).

\subsection{Extraction of Locality Principles and Generation of APP Scripts}

Briefly, we define the extraction of locality principles from a choreography by
attributing \cd{data} \cd{locality}---the principle that the closer the function
is to the data the lower its latency, proportional to faster access to the data
repository---to all functions that access a database (we omitted this
information from \cref{lst:choreo}, but it is simple to annotate \cd{services}
accordingly); \cd{call} \cd{locality} comes from interactions among functions,
in particular the repeated ones, which can benefit from running on machines with
faster access to the medium that accepts/delivers the call; \cd{code}
\cd{locality} comes from the re-use of loaded code in a worker's memory (i.e.,
avoiding fetching and loading times). On the right of \cref{fig:localities}, we
find an example of the extracted localities from \cref{lst:choreo}.

\begin{figure}[t]
\begin{center}
\noindent\begin{minipage}{.49\textwidth}
\begin{codebox}[.96\textwidth]
\begin{adjustbox}{width=.95\textwidth}
\begin{lstlisting}[mathescape=true]
# Extracted localities from choreo
data locality: 
  ( f, DB1 ), 
  ( f, DB2 ),
  ( h, DB3 ),
call locality: 
  ( f, g, SNS, 1:n ) # n = n. of labels/imgs
  ( g, h, SNS, 1:1 )
code locality: 
  ( g, h ) # Model
\end{lstlisting}
\end{adjustbox}
\end{codebox}
\end{minipage}
\begin{minipage}{.49\textwidth}
\begin{codebox}[\textwidth]
\begin{adjustbox}{width=.85\textwidth}
\begin{lstlisting}[mathescape=true]
# Infrastructure topology (speed)
( DB1, group1 ): 100
( DB2, group2 ): 80
( DB2, group1 ): 20
( DB3, group2 ): 50
( SNS, group1 ): 50
( SNS, group2 ): 50

# User-defined constraints 
anti-affine: ( f, g ), ( g, g ), ( h, g )
\end{lstlisting}
\end{adjustbox}
\end{codebox}
\end{minipage}
\end{center}
\caption{Left: locality principles extracted from the example from
\cref{lst:choreo}. Right: infrastructure topology and user-defined function
scheduling constraints.}
\label{fig:localities}
\end{figure}

The last ingredient is the infrastructure topology and the constraints that
users might want to impose on the functions. We report on the right of
\cref{fig:localities} an example of such a schema. In the example, the writing
\cd{( a, b ): N} indicates that \cd{a} and \cd{b} have a connection speed
(symmetric) of \cd{N} (we can abstract away the unit of measure, as long as all
items use the same), e.g., \cd{( DB1, group1 ): 100} means that the machines in
\cd{group1} have a (fast) connection of \cd{100} with \cd{DB1}. Note that the
absence of infrastructural pairs are as important as the present ones, e.g., the
fact that there is no couple \cd{( DB3, group1 )} in the schema means that no
machine in \cd{group1} can reach \cd{DB3}.

For compactness, we understand the specification of user-defined (anti-)affinity
constraints of the schema as symmetric, e.g., if \cd{(a, b)} are anti-affine, we
read this constraint as ``neither \cd{a} can run on a worker where a \cd{b} is
running nor \cd{b} can run on a worker where an \cd{a} is running''. Above, we
set \cd{f} and \cd{g} anti-affine to avoid running \cd{f} on a worker loaded
with \cd{g} (which performs heavy computations to train the model) and vice
versa. Similarly, we avoid placing more instances of the function \cd{g} on the
same worker and placing the functions \cd{g} and \cd{h} together. 

\begin{figure}
\begin{center}
\begin{adjustbox}{width=.9\textwidth}
\begin{minipage}{.28\textwidth}
\begin{codebox}[.9\textwidth]
\begin{lstlisting}[mathescape=true,basicstyle=\small\ttfamily]
f:
 - $\hlcode{workers}$:  group1
   $\hlcode{affinity}$: f, !g
 - $\hlcode{workers}$:  group1
   $\hlcode{affinity}$: !g
\end{lstlisting}
\end{codebox}
\end{minipage}
\begin{minipage}{.34\textwidth}
\begin{codebox}[.9\textwidth]
\begin{lstlisting}[mathescape=true,firstnumber=last,basicstyle=\small\ttfamily]
g: 
 - $\hlcode{workers}$: $\role{*}$
   $\hlcode{affinity}$: !f, !g, !h
\end{lstlisting}
\end{codebox}
\end{minipage}
\begin{minipage}{.27\textwidth}
\begin{codebox}[.94\textwidth]
\begin{lstlisting}[mathescape=true,firstnumber=last,basicstyle=\small\ttfamily]
h: 
 - $\hlcode{workers}$: group2
   $\hlcode{affinity}$: h, !g
 - $\hlcode{workers}$: group2
   $\hlcode{affinity}$: !g
\end{lstlisting}
\end{codebox}
\end{minipage}
\end{adjustbox}
\end{center}
\caption{APP script generated from the localities, topology, and user-defined constraints from \cref{fig:localities}}
\label{fig:app}
\end{figure}

Given the ingredients from \cref{fig:localities}, we can obtain the APP script
reported in \cref{fig:app}. In the script, we have two subsequent blocks for the
allocation of function \cd{f}. In both blocks, we try to allocate \cd{f} on
\cd{group1} because this is the only group of machines that can access both
\cd{DB1} and \cd{DB2}. Considering affinities, in the first block, we try to
allocate \cd{f} with other instances of the same function to exploit connection
pooling to \cd{DB1} and \cd{DB2}. Following the user-defined constraints, we set
a negative affinity with function \cd{g} (written \cd{!g}). In the second block,
\cd{f} can run on a worker without another instance of the same function running
on it (this item avoids the problem of self-affinity, which would prevent the
allocation of an initial \cd{f}), yet we preserve the anti-affinity with \cd{g}.

Since \cd{g} has no ``favourite'' group in the infrastructure (both \cd{group1}
and \cd{group2} reach the same speed w.r.t.\@ the only infrastructural locality
it presents, \cd{SNS}), the policy for \cd{g} allocates the function on any
available worker \role{*}. Following the anti-affinity constraints specified by
the user, we mark \cd{g} anti-affine with \cd{f} (as per the
symmetric interpretation of the anti-affinity constraints above), itself, and \cd{h} (similarly to the anti-affinity with \cd{f}).

Finally, we specify that function \cd{h} can only run on machines of \cd{group2}
since these are the only ones that can reach \cd{DB3}. Following the user-defined constraints, we have \cd{h} anti-affine with \cd{g} and \cd{h} affine with itself to exploit connection pooling.

Note that the synthesis of the APP script does not include all extracted
localities (cf.~\cref{fig:localities}). For instance, \cd{call} \cd{locality}
did not influence the script, due to the fact that \cd{group1} and \cd{group2}
have the same speed w.r.t.\@ \cd{SNS}. Another example is the \cd{code}
\cd{locality} between \cd{g} and \cd{h}, which share the same code dependency
(\cd{Model}, cf.\@ \cref{lst:choreo}) but which we could not indicate as affine
in the APP script due to the user-defined anti-affinity constraints (which have
priority over the extracted localities).

\section{Conclusion}

We illustrate \lang{}, a language proposal for exploring the
design space of applying CP to FaaS programming. Besides showcasing relevant
features of a CP language for FaaS, we provide application examples that target
projection and the management of the scheduling of functions.

In the future, we plan to conduct a formal investigation into the expressiveness
and limitations of \lang{}, considering other use cases taken from realistic
FaaS architectures and deepening the analysis of the interplay between
\cd{services}, \cd{stateful} participants, and \cd{stateless} functions. Another
interesting direction is to formally analyse the processes of extraction of
locality principles from choreographies and the mechanisation of the synthesis
of APP scripts.

\bibliographystyle{ACM-Reference-Format}
\bibliography{biblio}

\end{document}